\documentclass[%
 reprint,
 amsmath,amssymb,
 aps,
]{revtex4-2}

\usepackage{graphicx}
\usepackage{dcolumn}
\usepackage{bm}

\begin{document}

\preprint{APS/123-QED}

\title{Generation of a single ion large oscillator}

\author{Ryoichi Saito$^{1,2}$}
\email{r\_saito@ee.es.osaka-u.ac.jp}
\author{Takashi Mukaiyama$^{1,2}$}%
\email{muka@ee.es.osaka-u.ac.jp}
\affiliation{
$^1$Graduate school of Engineering Science, Osaka University, \\1-3 Machikaneyama, Toyonaka, Osaka 560-8531, Japan\\
$^2$Quantum Information and Quantum Biology Division, \\Institute for Open and Transdisciplinary Research Institute, Osaka University, Osaka 560-8351, Japan.\\
}%

\date{\today}

\begin{abstract}
We demonstrate the generation of a trapped ion oscillator having large oscillation amplitude of $16.9~{\rm \mu m}$.
Applying an offset voltage to the ion trap electrode helped achieve a displacement of the trap center within the time scale of 5 ns. The fluorescence dynamics of the ion were analyzed after the displacement to estimate the oscillation amplitude.
The realized trap displacement is one order magnitude larger than that achieved in the previous work.
Thus, this result is an important step toward the realization of a gyroscope using a single trapped ion.
\end{abstract}

\maketitle


Atoms in a coherent superposition of different momentum states enable high-precision measurement of physical quantities~\cite{PhysRevLett.67.181, geiger2011detecting, PhysRevA.74.023615, PhysRevLett.81.971, PhysRevA.65.033608, peters1999measurement}.
A remarkable example of this is an atom-based Sagnac interferometer~\cite{PhysRevLett.67.177, PhysRevLett.78.2046, PhysRevA.80.061603, PhysRevLett.99.173201}.
The short wavelength and low traveling velocity of the atoms cover the relatively small interference area, making the atom-based Sagnac interferometer comparable to other gyroscopes.

Recently, a scheme of rotation sensing using a single trapped ion was proposed by Campbell and Hamilton~\cite{Campbell_2017}.
Because an ion trap typically confines an ion tightly, realization of a large interference area in a Sagnac interferometer is challenging.
The study~\cite{Campbell_2017} proposes the use of spin-dependent momentum kicks to stimulate an ion into the spin-momentum entangled state and simultaneously accelerate the ion wave packets to realize a large interference area.
In addition, a fast displacement of the trap center on the scale of $100~{\rm \mu m}$ is used to further increase the interference area. The multiple momentum kicks have already been demonstrated by the NIST group~\cite{johnson2017ultrafast} up to the momentum of 200~$\hbar k$.
Therefore, the multiple accelerations of the ion is a practically feasible technique. In contrast, the fast displacement of the trap center has only been realized on the scale of $1.5~{\rm \mu m}$~\cite{alonso2016generation}, which is much smaller than the value expected in the proposal.
Since the size of the displacement primarily determines the interference area, the realization of large displacement is crucial for the operation of the gyroscope.

Here we experimentally fabricate a single trapped ion oscillator having a large oscillation amplitude using an ion at a Doppler temperature. We applied an extra voltage offset onto the ion trap electrode to realize a displacement of the trap center at a fast pace.
After applying the voltage for a certain amount of delay time, we brought the voltage back to the original value. If the delay time matches an integral multiple of the trap period, the ion oscillation motion vanishes completely;. however, if it does not match, a certain amount of the ion oscillation motion remains.
We analyzed the fluorescence dynamics of the ion after turning back the offset voltage, and precisely determined the amplitude of the ion oscillation motion. The maximum trap displacement that we were able to realize in this work is $16.9~\rm{\mu m}$, which is more than one order of magnitude larger than the results in a previous work~\cite{alonso2016generation}.
The coherence of the ion oscillation is measured to be over $1~{\rm ms}$. Thus, this result is an important step towards the realization of rotation sensing using a trapped ion. 

We trapped a single $\rm{Yb^+}$ ion within a conventional linear Paul trap, and the radial confinement of the trap was realized by an rf electric field at $24.7~\rm{MHz}$.
The axial confinement was generated by a static voltage applied to the electrodes installed at both sides of the trapping region in the axial direction.
The trap frequencies were $\left( \omega_{\rm r}, \omega_{\rm a} \right) = 2\times \pi \left( 1.83, 0.36 \right)~\rm{MHz}$, where $\omega_{\rm r}$ and $\omega_{\rm a}$ represent the radial and axial trap frequencies, respectively.
The trapped ion was cooled down to the Doppler cooling limit using the $S_{1/2}$-$P_{1/2}$ transition by a $369~\rm{nm}$ laser.
Further details of our experimental setup can be found here~\cite{doi:10.1063/5.0046121}.

To generate the oscillation of the ion, we superimposed the square-pulse voltage for a duration of $\tau$, on one side of static voltage for achieving the axial confinement.
The voltage pulse suddenly displaced the trap center of the ion by $x_{\rm d}$, and therefore the ion started oscillated with an amplitude of $x_{\rm d}$
in the trap for the duration that the voltage pulse was applied.
After the voltage pulse was turned off, the trap center returned to its initial position.
However, when the pulse duration $\tau$ matched the integral multiple of the trap period, the ion returned to the position of the initial trap center resulting in the ion oscillation vanishing suddenly. 
Otherwise, the ion maintained its oscillation depending on the phase of the ion oscillation at the instant when the voltage pulse was turned off.
The remaining ion oscillation provided the information of the extent of trap center displacement that was realized due to the voltage pulse.
The rise and fall time of the voltage pulse was approximately $5~\rm{ns}$, which is approximately $500$ times faster than the axial trap period in our experiment.

Furthermore, to detect the ion oscillation initiated by trap center displacement, we measured the time evolution of the ion fluorescence of the $S_{1/2}$-$P_{1/2}$ transition~\cite{biercuk2010ultrasensitive, PhysRevA.103.023105}.
We irradiate 369 nm laser with tens of MHz red detuned from the resonance to the ion along the axial direction for detection.
The scattering rate observed reflects the velocity of the ion in the oscillatory motion due to the Doppler shift.
Consequently, the fluorescence photons from the ion were collected by an objective lens and detected by a photomultiplier tube (PMT), and the number and arrival timing of the photon pulses from the PMT were recorded using a multi-channel scaler (MCS) within milliseconds.

\begin{figure}[tb]
\includegraphics[clip, width = 9.5cm]{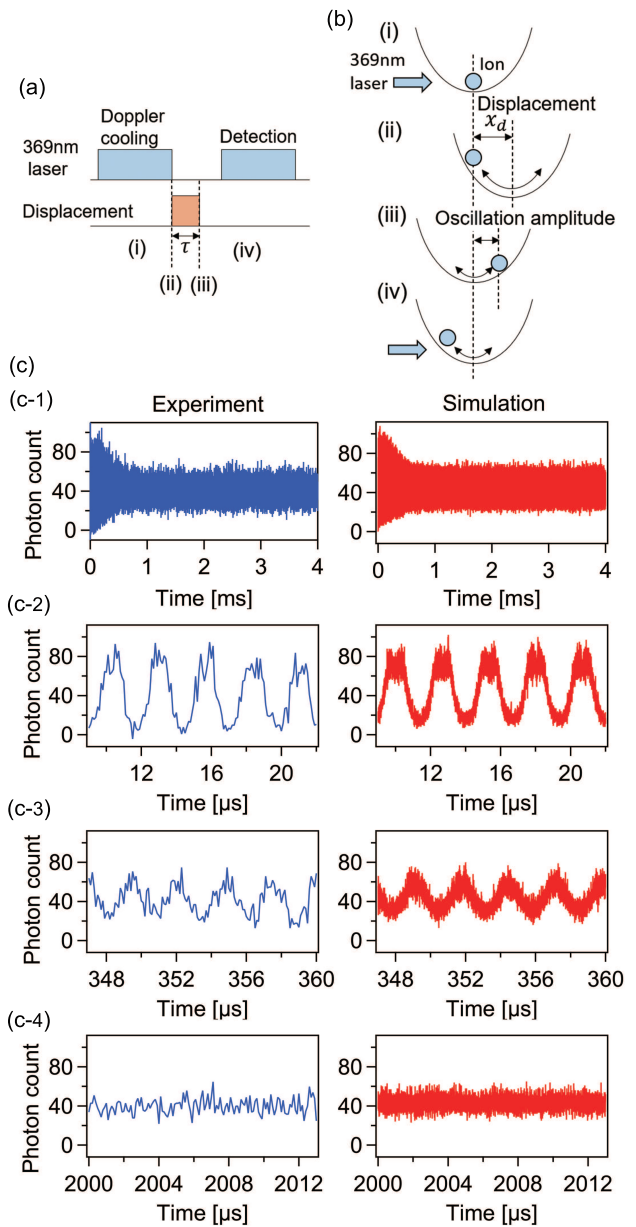}
\caption{\label{fig:fig1}
(a) Experimental sequence. (b) Schematic of trap displacement and ion oscillation.
(i) Doppler cooling for initialization, (ii) trap center displacement, (iii) returning of trap center displacement to the initial position,
(iv) detection of ion oscillation.
(c) Comparison of experimental and simulated ion fluorescence dynamics of overall dynamics up to $4~\rm{ms}$ (c-1),
9 to 22~$\rm{\mu s}$ (c-2), 347 to 360~$\rm{\mu s}$ (c-3), 2000 to 2013~$\rm{\mu s}$ (c-4), respectively.}
\end{figure}

The experimental sequence is illustrated in Figs. \ref{fig:fig1}(a), (b).
At first, (i) an ion is cooled near the Doppler cooling limit. 
Then, (ii), the cooling laser was shut off using an acousto-optics modulator (AOM). 
The square-pulse voltage rises instantaneously, and the trap center displaces in the axial direction.
After the pulse duration $\tau$, (iii) the pulse voltage reduced momentarily, and the trap center returned to its initial position.
Subsequently, (iv) $369~\rm{nm}$ laser for detection was turned on to detect the ion oscillation.
This sequence was repeated approximately 10000–30000 times to measure the oscillation amplitude.

Furthermore, a Monte Carlo simulation of an ion in a harmonic potential was performed to evaluate the experimental fluorescence dynamics.
The one-dimensional ion motion in a harmonic potential was calculated using the fourth-order Runge-Kutta method coupled with the stochastic scattering force for a model of photon absorption and emission by the detection laser. 

Figs. \ref{fig:fig1}(c) represent the typical experimental results and Monte Carlo simulation of the ion oscillation dynamics.
The horizontal axis depicts the elapsed time after turning on the detection laser.
The vertical axis indicates the arrival number of photons for each time period.
Certain points of the experimental data show negative photon count because we subtracted the averaged background level due to the scattering of light by the electrodes and stray light.
The displacement pulse time $\tau$ and amplitude of voltage pulse were $0.9~\rm{\mu s}$ and $0.1~\rm{V}$, respectively.
The frequency of the detection laser was set to $40~\rm{MHz}$ red detuned from the resonance.
Furthermore, the experimental results were obtained from an accumulation of $16715$ repeated sequences.

Fig. \ref{fig:fig1}(c-1) illustrates the overall fluorescence dynamics during the detection time of $4~\rm{ms}$.
In the simulation, oscillation amplitude of $8.5~\rm{\mu m}$ was chosen with a saturation parameter of the detection laser of $s = 15$.
The photon count decay from $0$–$0.5~\rm{ms}$ shows the decrease in oscillation amplitude due to the cooling-off motion of the detection laser, 
which is also reproduced by the simulation.
Figs. \ref{fig:fig1}(c-2, 3, 4) represent the enlarged graphs of Fig. \ref{fig:fig1}(c-1) along the horizontal axis.
Furthermore, the fluorescence dynamics immediately after irradiating the detection laser has been shown in Fig. \ref{fig:fig1}(c-2).
Fig. \ref{fig:fig1}(c-3) shows the fluorescence dynamics at approximately $350~\rm{\mu s}$.
The oscillation amplitude decays with time according to the cooling due to the detection laser, and
the signal amplitude decreases compared to the signal of Fig. \ref{fig:fig1}(c-2).
However, the generated oscillation due to the displacement manipulation is completely decayed after $1~\rm{ms}$ as shown in Fig. \ref{fig:fig1}(c-4).
Thus, the simulation was able to successfully reproduce the experimental result of the overall fluorescence dynamics, 
along with the time evolution of the ion fluorescence in a short time scale.

Moreover, the ion coherent motion and its velocity in the harmonic potential is a sine wave.
Therefore, the Doppler shift due to the oscillation can be expressed as $2\pi v \sin\left( \omega_{\rm{a}} t + \phi \right) / \lambda$.
Here, $t$, $\lambda$, $v$, and $\phi$ are the time, wavelength of the $369~\rm{nm}$ detection laser, velocity of the ion at the bottom of the harmonic potential, and arbitrary phase, respectively.
Furthermore, the spectrum of the $S_{1/2}$-$P_{1/2}$ transition is a Lorenz function.
Thus, the fluorescence dynamics $f \left( t \right)$ can be written as
\begin{equation}\label{dynamics}
	f \left( t \right) = \frac{s \Gamma/2}{1+s+4 \left\{ \delta - 2\pi v \sin \left(  \omega_{\rm a} t + \phi \right)/\lambda \right\}^2/\Gamma^2},
\end{equation}
where $\Gamma$, and $\delta$ are line width of $19.6~\rm{MHz}$ $S_{1/2}$-$P_{1/2}$ transition, and detuning of the detection laser.

The non-linearity of the spectrum shape causes the vertical asymmetry of the signal in Fig. \ref{fig:fig1}(c-2), as described by Eq.(\ref{dynamics}).
Moreover, when the Doppler shift exceeds the resonance of transition, the scattering rate decreases.
Therefore, the tops of the signal in Fig. \ref{fig:fig1}(c-2) attain a flat shape.
In the case of the small oscillation amplitude such as in Fig. \ref{fig:fig1}(c-3), 
the spectrum shape can be seen as a linear function around the detuning of the detection laser.
Thus, the fluorescence dynamics approach a sine function.

To estimate oscillation amplitude generated due to the displacement manipulation from the experimental results,
we fit the starting point of the fluorescence dynamics, which was not affected by the cooling dynamics, using Eq.(\ref{dynamics}).
The oscillation amplitude $x_{\rm{a}}$ can be expressed as $x_{\rm{a}} = v/\omega_{\rm{a}}$ and can be obtained using the fitting parameter $v$ and $\omega_{\rm{a}}$.
An estimated oscillation amplitude from the fitting of the experimental data in Figs. \ref{fig:fig1}(c) is $8.2\pm0.8~{\rm{\mu m}}$, which is consistent with the initial oscillation amplitude of $8.5~\rm{\mu m}$ of the simulation.

\begin{figure*}[tb]
\includegraphics[clip, width = 17cm]{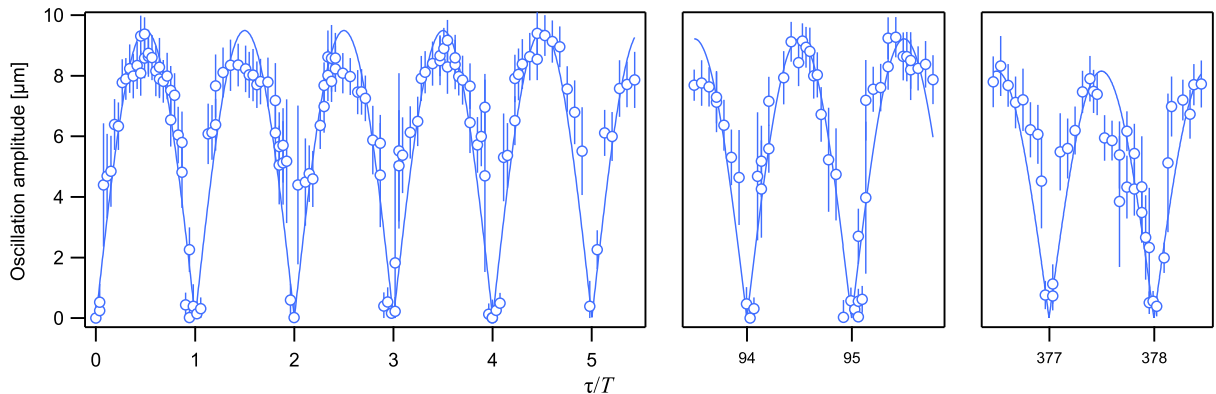}
\caption{\label{fig:fig2}
Pulse time dependence of the generated amplitude of an ion oscillator.
The vertical axis and the horizontal axis show the generated oscillation amplitude of the ion after the displacement
and the normalized pulse time by the axial trap period, respectively.
Each point represents measured data.
The solid line is the fitted line of the points obtained via Eq.(\ref{bang_bang}), and the realized trap displacement $x_{\rm d}$ is $4.7\pm0.1~\rm{\mu m}$, which is obtained by the fitting of the left side column data.
}
\end{figure*}

To analyze the relation between the ion displacement size and the displacement pulse time $\tau$, we measured the ion oscillation amplitude as a function of the pulse duration $\tau$ normalized by the trap period $T$. The results are shown in Fig. \ref{fig:fig2}.
The dips within the oscillation amplitude are observed in every integral multiple of the axial trap period.
When $\tau / T$ attains an integer value, the ion position matches the initial trap center with a zero velocity of oscillation.
Thus, the ion returns back to the bottom of the harmonic potential without oscillating.
Although $\tau / T$ attains a half-integer value, the remaining ion oscillation is maximized and the amplitude attains a value which is twice the trap center displacement $x_{\rm d}$ value.

Thus, we fit the measured oscillation amplitudes using equation\cite{alonso2016generation} to estimate the generated trap displacement $x_{\rm d}$ due to the applied square-pulse voltage,
\begin{equation}\label{bang_bang}
	x_{\rm a} = x_{\rm d} \sqrt{2\left\{ 1-\cos\left( 2\pi \tau/T \right)\right\} }.
\end{equation}
Thus, the obtained displacement, $x_{\rm d}$, is $4.7\pm0.1~\rm{\mu m}$ after fitting of Eq.(\ref{bang_bang}).
An ion that is cooled near the Doppler cooling limit in the harmonic potential localizes at the sub-micrometer region.
Thus, this displacement size is large compared to the ion localized area, and applied square-pulse voltage can generate a large ion mechanical oscillator.

The pulse time dependence of the oscillation at $\tau/T = 95$ and $378$ are also shown in Fig.\ref{fig:fig2}, to evaluate the performance of long time trap displacement.
$\tau/T = 95$ and $378$ correspond to $267~\rm{\mu s}$ and $1068~\rm{\mu s}$ time evolution, respectively, in our experiment.
The obtained $x_{\rm d}$ from the fitting of Eq.(\ref{bang_bang}) is $4.6\pm0.1~\rm{\mu m}$ using the result of the middle column in Fig. \ref{fig:fig2}.
Therefore, the trap displacement during $95$ oscillations was performed at a quality level equivalent to the trap displacement in a short time.
However, in the case of $\tau /T = 378$ and other longer time evolutions, the contrast within the results decrease.
This decreasing oscillation amplitude can be considered as being caused primarily by the trap frequency drift during the measurement.
The trap frequency drift of $0.3~\%$ induces more than one period shift in the case of $\tau /T = 378$.
Therefore, the axial trap frequency needs to be stabilized for precise control of a long-time trap displacement operation.
Moreover, the imperfection of the dips may be attained by the heating of the ion.

\begin{figure}[tb]
\includegraphics[clip, width = 8.5cm]{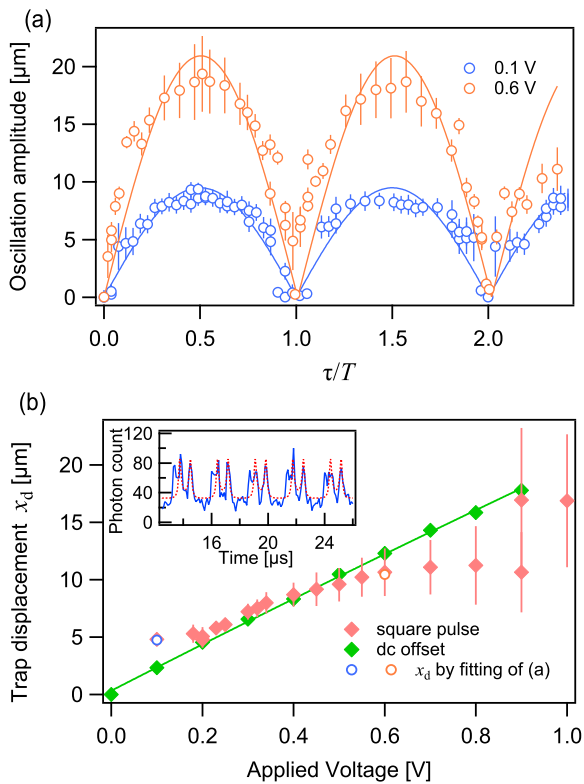}
\caption{\label{fig:fig3}
(a)Pulse time dependence of an ion oscillator size.
The measured oscillation amplitude of an ion by the fluorescence dynamics is represented as a function of the applied square-pulse time for the trap center displacement. 
The open dark blue and orange circles show the square-pulse with amplitude of $0.1~{\rm V}$ and $0.6~{\rm V}$, respectively.
The solid curve is the fitting line by Eq.(\ref{bang_bang}).
(b)Generated trap displacement vs. applied voltage to the electrode for the axial confinement.
The filled red and blue diamonds represent the measured displacement due to the square-pulse and the dc offset voltage, respectively.
The displacements by the square-pulse are evaluated by the half value of the oscillation amplitude under the condition of $\tau = T/2$.
In the case of applying the dc offset voltage, we measure the trapping position shift by the fluorescence image of the ion.
The trap displacements estimated by (a) are shown as open circles.
(inset) Starting point of the fluorescence dynamics with the square-pulse of $0.4~{\rm V}$ amplitude. The Blue solid line and red dotted line show experimental results and fitting by Eq.(\ref{dynamics}).
}
\end{figure}

To generate a larger ion mechanical oscillator,
we raised the voltage of the square-pulse for the trap displacement.
The pulse time dependence of oscillation amplitude with the pulse voltage of $0.1~\rm{V}$ and $0.6~\rm{V}$ is shown in Fig.\ref{fig:fig3}(a).
Each point in the figure represents the data measured by the fluorescence dynamics.
The solid curve represents the fitting line by Eq.(\ref{bang_bang}).
The plot of $0.1~\rm{V}$ square-pulse represents the same data in Fig.\ref{fig:fig2}.

The oscillator size generated by $0.6~\rm{V}$ square-pulse is greater than twice the size in the case of $0.1~\rm{V}$.
The trap displacement $x_{\rm{d}}$ estimated by fitting of Eq.(\ref{bang_bang}) is $10.5\pm0.4~\rm{\mu m}$, and the plot data is reasonably consistent with the fitting line of Eq.(\ref{bang_bang}).
Therefore, we consider trap displacement manipulated as expected when the instantaneous pulse voltage, with amplitude as $20~\%$ of the static voltage of axial confinement, is applied to the electrode.

Subsequently, to evaluate the consistency of the trap displacement manipulation, we measured the trap displacement by applying the square-pulse, and a dc offset to the electrode for the axial confinement, respectively. 
This comparison is represented in Fig.\ref{fig:fig3}(b), which indicates that the trap displacements are either a function of the applied square-pulse voltage or is a dc offset voltage.

The plots with the filled green diamonds indicate the measured trap displacement due to changing of the dc offset voltage.
The trap displacement is observed by the shift in position of the trapping ion using the fluorescence image of an electron-multiplying CCD camera.
We fit the data using the square root function of the voltage shown as by the solid green curve in Fig.\ref{fig:fig3}(b).

In contrast, the plots with the filled red diamonds show the trap displacement due to square-pulse voltage.
We fixed the pulse time at $\tau = T/2$ and measured the oscillation amplitudes at each pulse voltage using the fluorescence dynamics to evaluate the trap displacements.
If the pulse time $\tau$ matched the half-integral multiple of the trap period, the generated oscillation amplitude attained a value of $2x_{\rm{d}}$ due to the displacement square-pulse manipulation.
Therefore, we plotted the half value of the estimated oscillation amplitude obtained from fluorescence dynamics in terms of the square-pulse as shown in Fig.\ref{fig:fig3}(b).
Further, we also represented the trap displacement $x_{\rm{d}}$ obtained from the results of Fig.\ref{fig:fig3}(a) as an open circle.
The starting point of the fluorescence dynamics with the square-pulse of $0.4~{\rm V}$ is shown as the inset of Fig.\ref{fig:fig3}(b).
The solid blue line and the dotted red line indicate the measured dynamics and fitted curve by Eq. (\ref{dynamics}), respectively.
The detection laser is detuned to $70~{\rm MHz}$ and the accumulated number to measure this data is 18867.

The trap displacements using the square-pulse correspond reasonably with those by the dc offset voltage.
This result is the obvious proof that a large size ion oscillator is generated upon using the square-pulse manipulation.
Thus, we generated the ion oscillator of size $16.9\pm5.8~{\rm{\mu m}}$ at most.
Furthermore, the values of the small applied voltage match well.
However, the deviation between the square-pulse and the dc offset becomes large with increasing applied voltage, and the error bars of the measured trap displacement due to the square-pulse also increase.
With the increasing size of the ion oscillator and the maximum Doppler shift due to the oscillation, the fluorescence dynamics signal is strongly modulated, and the double-peak shape (shown in Figs.\ref{fig:fig1}(c-2)) becomes more pronounced(shown in the inset of Fig.\ref{fig:fig3}(b)).
Herein, a left-right asymmetry of the double peak shape was observed.
This asymmetry was considered to be caused by the misalignment of the detection laser
because the ion becomes sensitive to laser power density, leading to inhomogeneity on the left and right of the trap center, with increasing oscillator size.
This asymmetry of fluorescence dynamics signal causes the uncertainty in the trap displacement size.

Furthermore, under this large displacement condition, the overall fluorescence dynamics shown in Fig.~\ref{fig:fig1}(c-1) suffers from further deformation and the Monte-Carlo simulation fails to reproduce the overall fluorescence dynamics.
To clarify the origin of the deformation, we investigate correlation between the displacement size and the radial confinement strength.
If we choose weaker radial confinement, the deformation of the fluorescence dynamics is more pronounced. Since the weaker radial confinement causes more radial-axial motional coupling, the result indicates that the deformation of the fluorescence dynamics arises form an anharmonicity of the trap potential which is the origin of the cross-directional coupling.

Finally, we discuss the performance of ion Sagnac interferometer in a two-dimensional orbit generated by the SDKs and the trap displacement.
The sensitivity of the ion gyroscope can be described as \( S = 1/(2N_{\rm k}\hbar k x_{\rm d} \sqrt{\Delta t}) \),
where $N_{\rm k}$, $\hbar$, $k$, and $\Delta t$ are the number of SDKs, reduced Planck constant, wavenumber of the laser for SDKs, and total interference time, respectively\cite{Campbell_2017}.
Assuming the number of SDKs $N_{\rm k} = 100$ \cite{johnson2017ultrafast}, the trap displacement achieved in this work $x_{\rm d}=16.9 {\rm \mu m}$, 
and $\Delta t = 1~{\rm ms}$ \cite{decoherence} yields $S \approx 1.8~{\rm deg/\sqrt{hour}} $. 
Repeating the measurement 30000 times provides sensitivity that is comparable to the typical ring laser gyroscope\cite{RevModPhys.57.61}.

In conclusion, we demonstrate the generation of a large ion mechanical oscillator via the instantaneous trap displacement achieved by applying the square-pulse voltage to the ion trap electrode.
The oscillator size is estimated by the newly developed scheme that is based on measuring the fluorescence dynamics.
The kinetic energy of the oscillation after the trap displacement manipulation vanishes when the square-pulse time matches the integral multiple of the trap period.
The time evolution during the trap displacement reaches values greater than $1~{\rm ms}$ when the oscillation of the ion is under control.
Moreover, the maximum trap displacement reached $16.9\pm5.8~\rm{\mu m}$ in our experiment.
This value is more than one order of magnitude larger than that observed in the previous work~\cite{alonso2016generation}.
Furthermore, the large displacement has been applied and confirmed using an ion in a Doppler temperature regime from the measurement of the fluorescence dynamics of the ion.
The present work makes an important step toward realizing a circular-orbit matter-wave interferometer in combination with the spin-motion entanglement.
A sensitivity of rotation sensing is expected to be $S\approx1.8~{\rm deg/\sqrt{hour}}$ with the trap displacement performed in this article.

This work is supported by JST-Mirai Program Grant Number JPMJMI17A3, Japan.


\bibliography{bib}

\end{document}